\documentclass[prd,amsmath,amssymb,floatfix]{revtex4}

\usepackage{graphicx}

\begin{document}

\newcommand{\vp}{\varphi}
\newcommand{\pd}{{\partial}}

\title{Anistropic Stars: Exact Solutions and Stability}

\author{Krsna Dev}
\email{kdev@haverford.edu}
\affiliation{Department of Physics,
         Haverford College
				 Haverford, PA 19041, USA}

\author{Marcelo Gleiser
\footnote{Invited contribution to the International Workshop on
Astronomy and Relativistic Astrophysics, October 12-16 2003, Olinda, Brazil.}}
\email{gleiser@dartmouth.edu}
\affiliation{Department of Physics and Astronomy,
         Dartmouth College
				 Hanover, NH 03755, USA}

\begin{abstract}
I report on recent work concerning the existence and
stability of self-gravitating spheres with anisotropic pressure. After
presenting new exact solutions, Chandrasekhar's variational
formalism for radial perturbations
is generalized to anisotropic objects and 
applied to investigate their stability. It is shown that
anisotropy can not only support stars of mass
$M$ and radius $R$ with $2M/R \geq 8/9$ and
arbitrarily large surface redshifts, but that stable
configurations exist for values of the adiabatic index
$\gamma$ smaller than the corresponding isotropic value.

\end{abstract}

\maketitle

\section{Introduction}
A common assumption in the study of stellar structure 
and evolution is  
that the interior of a star can be modeled as a perfect 
fluid \cite{CLAYTON}.  This
perfect fluid model  necessarily requires the pressure in the interior 
of a star to be isotropic. This approach  has been used extensively in the 
study of polytropes, including white dwarfs, and of compact objects 
such as neutron stars \cite{GLENDENNING}.
However, theoretical advances in the last
decades indicate that deviations from local isotropy in the 
pressure, in particular at very high densities, may play an 
important role in determining stellar properties \cite{RUDERMAN}. 

The physical situations where anisotropic pressure may be relevant
are very diverse. By anisotropic pressure we mean that the radial component
of the pressure, $p_r(r)$, differs from the angular components, $p_{\theta}(r)
=p_{\varphi}(r)\equiv p_t(r)$. (That $p_{\theta}(r)=p_{\varphi}(r)$  is a
direct consequence of spherical symmetry.) Of
course, spherical symmetry demands both to be strictly a function of the radial
coordinate.
Boson stars, 
hypothetical self-gravitating compact objects resulting from the 
coupling of a complex scalar 
field to gravity, are systems where anisotropic pressure occurs
naturally \cite{B-STARS}. In the interior of neutron stars pions may condense. 
It has been shown that due to the 
geometry of the ${\pi}^{-}$ modes, anisotropic distributions of pressure 
could be 
considered to describe a pion condensed phase configuration \cite{SAWYER}. The
existence of solid cores and type P superfluidity
may also lead to departures from
isotropy within the neutron star interior \cite{GLENDENNING}. 

Since
we still do not have a detailed microscopic formulation of the possible
anisotropic stresses emerging in these and other
contexts, we take the general approach 
of finding several exact solutions representing different physical
situations, modeled by {\it ansatze} for the anisotropy factor,
$p_t-p_r$.  Previous studies have
found some exact solutions, assuming certain relations for the anisotropy 
factor \cite{BOWERS}.
Our goal here is two fold: first, to find new
exact solutions which may better model realistic situations
and explore their physical properties \cite{kg1}; second, to investigate their
stability against small radial perturbations \cite{kg2}. For this, we generalize
Chandrasekhar's well-known variational approach to anisotropic
objects. We find that not only interesting exact solutions can be
found \cite{kg1}, but that they may have a wider stability range when
contrasted with their isotropic counterparts \cite{kg2}.

\section{Relativistic Self-Gravitating Spheres: Basics}

We consider a static equilibrium distribution of matter 
which is spherically
symmetric. In Schwarzschild coordinates the metric can be written as 

\begin{equation}
ds^2 = e^{\nu}dt^2  - e^{\lambda}dr^2 - r{^2}d{\theta}^2 - 
r{^2}\sin^{2}{\theta}d{\phi}^2~,
\end{equation}
where all functions depend only on the radial coordinate $r$.
\noindent The most general energy-momentum tensor compatible with spherical
symmetry is
\begin{equation}
T^{\mu}_{\nu} = {\rm diag}(\rho, -p_{r}, -p_{t}, -p_{t})~.
\end{equation}
We see that isotropy is not required by spherical symmetry; it is an added
condition. 
The Einstein field equations for this spacetime geometry and matter 
distribution are
\begin{equation}
\label{einstein.1}
e^{-\lambda} \left( \frac{{\nu}^{\prime}}{r}+ \frac{1}{r^2} \right)  - 
\frac{1}{r^2} = 8{\pi}p_{r}~;
\end{equation}
\begin{equation}
\label{einstein.2}
e^{-\lambda} \left( \frac{1}{2} {\nu}^{\prime \prime} - 
\frac{1}{4}{\lambda}^{\prime}{\nu}^{\prime} 
+ \frac{1}{4} \left({\nu}^{\prime} \right)^{2} +  \frac{\left({\nu}^{\prime} - 
{\lambda}^{\prime} \right)}{2r} \right) = 8{\pi}p_{t}~;
\end{equation}

\begin{equation}
\label{einstein.3}
e^{\lambda} \left( \frac{{\lambda}^{\prime}}{r} - \frac{1}{r^2} \right ) + 
\frac{1}{r^2} = 8{\pi}{\rho}~.
\end{equation}
\noindent Note that 
this is a system of 3 equations with 5 unknowns. Consequently, it is 
necessary to specify two equations of state, such as ${p_{r} = p_{r}(\rho)}$ 
and ${p_{t} = p_{t}(\rho)}$.

\noindent It is often
useful to transform the above equations into a form where the 
hydrodynamical properties of the system are more evident. 
For systems with isotropic 
pressure, this formulation results in the 
Tolman-Oppenheimer-Volkov (TOV) equation. 
The  generalized TOV equation, including
anisotropy, is 
\begin{equation}
\frac{dp_{r}}{dr}  = -(\rho + p_{r})\frac{{\nu}^{\prime}}{2} + 
\frac{2}{r}(p_{t} - p_{r})~,
\end{equation}
\noindent   with 
\begin{equation}
\frac{1}{2}{\nu}^{\prime} = \frac{m(r) + 4 \pi r^{3} p_{r}}{r(r - 2m)},
\end{equation}
\noindent and 
\begin{equation}
m(r) = \int_{0}^{r} 4 \pi r^{2} \rho dr~.
\end{equation}
\noindent Taking $r = R$ in the above expression gives us the
Schwarzschild mass, $M$. [This implicitly assumes that $\rho=0$ for $r>R$.] 

In order to solve the above equations we must impose appropriate boundary 
conditions. We require that the solution be regular at the origin. This 
imposes the condition that $m(r) \rightarrow 0$ as $r \rightarrow 0$. If 
$p_{r}$ is finite at the origin then ${\nu}^{\prime} \rightarrow 0$
as $ r \rightarrow 0$. The gradient $dp_{r}/dr$ will be finite at 
$r =0$ if $(p_{t} - p_{r})$ vanishes at least as 
rapidly as $r$ when $r \rightarrow 0$. This will be the case in all scenarios
examined here. 

The radius of the star is determined by the condition $p_{r}(R) =0$. It is 
not necessary for $p_{t}(R)$ to vanish at the surface. But it is reasonable to 
assume that all physically interesting solutions will have $p_{r} ,p_{t} \geq 
0 $ for $r \leq R$.

\section{Exact Solutions}

In ref. \cite{kg1}, we obtained several exact solutions for different forms of the
pressure anisotropy. Our solutions fall into two classes: i) $\rho$ = constant,
and ii) $\rho \propto 1/r^2$. Given the limited space, we will restrict
the presentation to the latter case. Interested readers should consult ref. \cite{kg1}
for details.

Consider stars with energy density modeled as
\begin{equation}
\label{density}
\rho = \frac{1}{8 \pi} \left( \frac{a}{r^{2}} + 3b \right),
\end{equation}
\noindent where both  $a$ and $b$ are constant. The choice of the values for 
$a$ and $b$ is dictated by the physical configuration under consideration.
For example, $ a = 3/7$ and $b = 0$, corresponds to a relativistic Fermi gas,
as in the Misner-Zapolsky solution for ultradense cores of neutron stars\cite{ZAPOLSKY}.
If we take $ a= 3/7$ and $b \neq 0$ then we have a relativistic Fermi gas
core immersed in a constant density background. For large $r$ the constant
density term dominates ($r_c^2\gg a/3b$), 
and can be thought of as modeling a shell 
surrounding the core. We also take the pressure anisotropy to be
\begin{equation}
\label{ansio}
p_{t}  - p_{r} = \frac{1}{8 \pi}\left( \frac{c}{r^{2}} + d \right)~,
\end{equation}
with $c$ and $d$ constant.

We found it convenient to seek
solutions for the metric function $ \nu(r) $ directly, rather than 
solving the generalized TOV 
equation. We then use the known functions $ \lambda(r) $ and $ \nu(r) $ 
to find the radial and tangential pressures. From eqs. \ref{einstein.1},
\ref{einstein.2}, and \ref{einstein.3}, 
we find 
\begin{equation}
\label{inveq}
\left( \frac{{\nu}^{\prime \prime}}{2} + \frac{ ({\nu}^{\prime })^{2}}{4} 
\right)
e^{-\lambda} - {\nu}^{\prime}\left( \frac{{\lambda}^{\prime}}{4} + 
\frac{1}{2r}\right)
e^{-\lambda}  - \left( \frac{1}{r^{2}} + 
\frac{ {\lambda}^{\prime}}{2r}\right)e^{-\lambda} 
+ \frac{1}{r^{2}} = 8 \pi(p_{t} - p_{r}).
\end{equation}
Introducing a new variable $y = e^{\frac{\nu}{2}}$, 
eq. \ref{inveq} becomes,
\begin{equation}
\label{yyy}
( y^{\prime \prime} )
e^{-\lambda} - {y}^{\prime}\left( \frac{{\lambda}^{\prime}}{2} + 
\frac{1}{r}\right)
e^{-\lambda}  -  y \left[( \frac{1}{r^{2}} + 
\frac{ {\lambda}^{\prime}}{2r})e^{-\lambda} 
- \frac{1}{r^{2}} \right] = 8 \pi y(p_{t} - p_{r}). 
\end{equation}
Since $ e^{-\lambda} = 1  - 2m(r)/r$, using eq. \ref{density}
we find
\begin{equation}
e^{-\lambda} = 1 - a - br^{2} \equiv I_{b}^{2}(r),
\end{equation}
where we defined the function $I_{b}^{2}(x) \equiv 1 - a -bx^{2}$.
When $b = 0$, we write $I_{0}^{2} \equiv 1 - a$.
Using the expression for $e^{-\lambda}$ in eq. \ref{yyy} and substituting 
for the pressure anisotropy we find
\begin{equation}
\label{yy2}
\left[br^{4} - (1-a)r^{2} \right]y^{\prime \prime} + (1-a)r y^{\prime} 
-(a -c -dr^{2})y = 0~.
\end{equation} 
The full solution of eq. \ref{yy2} with $ a,b,c,d \neq 0$ is in \cite{kg1}.
Here, we will only consider solutions with $b=d=0$.
In this case, the total mass is $M=aR/2$, and $\exp[-\lambda] = 1-a$.
Since for any static spherically-symmetric configuration we expect 
$(2M/R)_{crit} \le 1$,  we must have $ a < 1$. (Also,
 the metric coefficient $g_{rr} $ becomes infinite when $a = 1$).

Since we want to construct stars with finite radii and density 
in the context of anisotropic pressure,
we impose boundary conditions such that $p_r(R)=0$. 
With $ b = d=0$,  eq. \ref{yy2}  reduces to 
 an Euler-Cauchy equation,
\begin{equation}
(1- a)r^{2} y^{\prime \prime} - (1 - a)r{y}^{\prime}  + (a - c )y =0. 
\end{equation}

The solutions of this equation fall into three classes, depending on the 
value of $q$ ($q$ is real, $q=0$, or $q$ is imaginary), where
\begin{equation}
q \equiv  \frac{( 1 +c - 2a)^{\frac{1}{2}}}{(1 - a)^{\frac{1}{2}}}.
\end{equation}
We will only show results for $q$ real. (See ref. \cite{kg1} for the other cases.)
 
The solution for $y$ is 
\begin{equation}
y = A_{+} \left(\frac{r}{R} \right)^{1+q} + 
A_{-} \left(\frac{r}{R} \right)^{1-q}~,
\end{equation}
with the constants $A_{+}$ and $A_{-}$ fixed by boundary conditions. 
For the case under consideration here ($ b = d=0$), the 
boundary conditions are
\begin{equation}
e^{-\lambda(R)} = e^{\nu(R)}  =  I_{0}^{2}, ~~~{\rm and}
~e^{\nu (R)} \frac{d \nu}{dr}|_R = \frac{a}{R}~.
\end{equation}
 Applying the boundary conditions we find 
\begin{equation}
A_{+} = \frac{I_{0}}{2} + \frac{1 - 3I_{0}^{2}}{4qI_0} ~~~~~ {\rm and} 
~~~~~A_{-} = A_{+}(q \rightarrow - q).
\end{equation}

\noindent The radial pressure for this case, after substituting the expressions
for $A_{+}$ and $A_{-}$, is 
\begin{equation}
8 \pi p_{r} = \frac{(3I_{0}^{2} - 1)^{2} - 4q^{2}I_{0}^{4}}{r^{2}}
\left[ \frac{R^{2q} - r^{2q}}{(3I_0^{2} - 1 + 2qI_0^{2})R^{2q}  
+ ( 1 -3I_0^{2} + 2qI_0^{2})r^{2q}} \right]~.
\end{equation}
\noindent We note  that the boundary conditions automatically guarantee that 
$p_r(R)=0$. 
The radial pressure is always greater than zero provided  $ a < 2/3 $ and 
$ a^{2} > 4c(1 - a)$. Since by definition $ a > 0$,
the second condition implies  $c > 0$. Thus, this model does not allow for 
negative 
anisotropy. Further, since we are considering the case $q > 0$, we must impose
the condition $ 1 + c < 2a $.
Combining the two inequalities for $a$ and $c$, we obtain, $ 2a -1 <
c < a^{2}/4(1-a) $. Since we have $ 0 < a < 2/3$ we find that $0 < c < 1/3$.
We note that for the anisotropic case the maximum value of 
$a $ is $ 2/3$, corresponding to a $33 $\% increase
when compared with the isotropic case ($a=3/7$). 
In figure 1 we plot the radial pressure, $p_{r}$, as a function of 
the radial coordinate $r$, 
for $a = 3/7$ and several values of $c$. Note that for this choice of $a$,
the inequality $c < a^2/4(1-a)$ imposes that $c<0.08$ for positive pressure
solutions. This can be seen in the figure. For larger anisotropies, no static 
self-gravitating stable configuration is possible. 
For $ r \ll R$,  we find
\begin{equation}
8 \pi r^2 p_{r} = 3I_0^{2} - 1 - 2qI_0^{2}~.
\end{equation} 

\noindent Choosing $a=3/7$ we recover, in the limit $c\rightarrow 0$, the 
Misner-Zapolsky solution \cite{ZAPOLSKY}, with 
$ p_{r} = 1/(56 \pi r^{2})=\rho/3$.

\begin{figure}
\hspace{0.75 in}
\includegraphics[width=3in,height=3in]{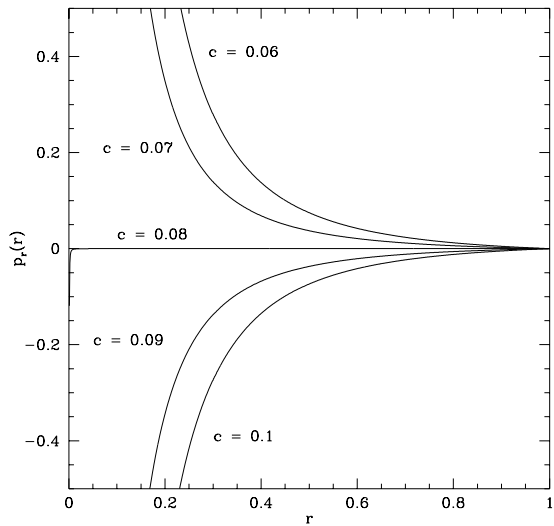}
   \caption{Radial pressure as a function of $r$ for $\rho$ and
            $(p_{t} - p_{r}) \propto r^{-2}$ and $q $ real.}
	 \label {pressure} 
\end{figure}

\section{Stability}

For sake of brevity, we will skip most details of our generalization
of Chandrasekhar's formalism to anisotropic spheres. Readers can consult
ref. \cite{kg2} for details. As in Chandrasekhar's original formalism, we
limit our study to small radial and baryon-number conserving perturbations. 
Writing $ \rho ~ = ~ {\rho}_{o} ~ + ~\delta \rho,
~ p_{r} ~ = ~ {p_{r}}_{o} ~+ ~\delta p_{r},
~ p_{t} ~ = ~ {p_{t}}_{o} ~+ ~\delta p_{t},
~\lambda ~ = ~ {\lambda}_{o} ~ + ~\delta \lambda,
$ and $~\nu ~ = ~ {\nu}_{o} ~ + ~\delta \nu $
we find that, to first order in $v=d\xi/dt$ ($\xi$ is
the Lagrangian displacement) and using the zeroth-order equations,
the perturbation in the radial pressure satisfies,
\begin{equation}
\delta p_{r} = - {p_{r_{o}}}^{\prime} - \gamma {p_{r}}_{o} 
\frac{e^{\nu_{o}/2}}{r^{2}} \left( r^{2} e^{ {\nu}_{o}/2} \xi \right)^{\prime}
+ \frac{2 \xi}{r} \Pi_{o} \frac{\partial p_{r o}}{\partial \rho_{o}}~,
\end{equation}
\noindent with $\gamma$ being the adiabatic exponent defined as
\begin{equation}
\gamma \equiv \frac{1}{p_{r} (\partial n/\partial p_{r})} \left[ n - (\rho
+ p_{r})\frac{\partial n}{\partial p_{r}} \right]~,
\end{equation}
\noindent and $\Pi\equiv p_t - p_r$.
We now assume that all perturbations have a time dependence of the form
$e^{i \omega t}$. Further, considering $\delta \lambda$, $ \delta \nu $,
$\delta \rho$, $\delta p_{r}$ and $\delta \Pi$ to now represent the amplitude
of the various perturbations with the same  time dependence we obtain,
after using the zeroth-order equations, the pulsation equation
governing the radial stability of anisotropic stars \cite{kg2}
\begin{eqnarray}
\label{iro24}
\lefteqn{\omega^{2}\left( \rho_{o} + {p_{r}}_{o} \right) \xi 
e^{ \lambda_{o} - \nu_{o}}
 = \frac{4}{r}{p_{r}}_{o}^{\prime} \xi
- e^{-(\lambda_{o} + 2 \nu_{o})/2} 
\left[ e^{(\lambda_{o} + 3 \nu_{o})/2} \gamma \frac{{p_{r}}_{o}}{r^{2}} 
(r^{2}e^{-\nu_{o}/2} \xi)^{\prime} \right]^{\prime} } +  \\  \nonumber
& & 8 \pi e^{\lambda_{o}}(\Pi_{o} 
+ {p_{r}}_{o})( \rho_{o} + {p_{r}}_{o}) \xi  
- \frac{1}{ (\rho_{o} + {p_{r}}_{o})} ({p_{r}}_{o}^{\prime})^{2} \xi 
 +  \frac{4 {{p_{r}}_{o}}^{\prime} \Pi_{o} \xi}{r( \rho_{o} + {p_{r}}_{o})}
-\frac{4 \Pi_{o}^{2}  \xi}{r^{2}( \rho_{o} + {p_{r}}_{o})} \\ \nonumber
& & - e^{-(\lambda_{o} + 2 \nu_{o})/2} 
\left[ e^{(\lambda_{o} + 2 \nu_{o})/2}\frac{2}{r} \xi \Pi_{o} 
\left(\frac{\partial p_{r}}{\partial \rho} + 1\right)  \right]^{\prime}   
 - \frac{8}{r^{2}}\Pi_{o}\xi 
 - \frac{2}{r} \delta \Pi ~.
\end{eqnarray}
\noindent The boundary conditions imposed on this equation are
\begin{equation}
\label{iro25}
\xi = 0 {\rm ~at ~} r = 0 {\rm ~~~and~~~} \delta p_{r} = 0 {\rm ~at~} r = R.
\end{equation}
 
The pulsation eq. (\ref{iro24}), together with the boundary conditions
eq. (\ref{iro25}), reduce to an eigenvalue problem for the 
frequency $\omega$ and
amplitude $\xi$. One multiplies this equation by 
$r^{2} \xi e^{(\lambda + \nu)/2}$ and integrates
over the entire range of $r$, using the orthogonality condition
\begin{equation}
 \int_{0}^{R}   e^{ (3\lambda - \nu)/2} 
\left( \rho + p_{r} \right) r^{2} {\xi}^{i} \xi^{j} dr = 0 
~~~~~~~~~(i \neq j)~,
\end{equation}
\noindent where $ \xi^{i}$ and $\xi^{j}$ are the proper solutions belonging
to different eigenvalues  $\omega^{2}$.

We now apply this equation to the exact solutions with $\rho\propto 1/r^2$
described above. Using the trial function
$\xi  = r^{2} ( \rho + p_{r}) e^{\nu}$ we found that
all integrals could be computed exactly.
In table 1
we present results for the frequencies of radial oscillations 
 ${\omega}^{2}$ as a function of the anisotropy parameter, $c$, for 
given values of the density parameter. Instabilities
set in for $\omega^2 < 0$. This can be used to find the
critical value for the adiabatic index, $\gamma_c(c)$,
and the maximum value for the anisotropy paramenter, $c_{max}$.
We also give, in table 2, the values of
$\gamma_{c}$ above  which stable oscillations are possible. Here we
see that the effect of a positive anisotropy is to reduce the value
of $\gamma$, thus giving rise to a more stable configuration when compared
with the corresponding isotropic model. In particular, for the Misner-Zapolsky
solution ($a=3/7$), we find that a small positive pressure anisotropy in the equation
of state improves the neutron star's core stability.

\vspace{0.5cm}

\begin{table}
\begin{center}
\begin{tabular}{|l|l|r|}  \hline

a =  2/9  & $\omega^{2}R^{2}$ =   0.95($\gamma$ -1.79) 
+ (101.1 -52.6  $\gamma$ )c \\  \hline

a = 2/7  &  $ \omega^{2}R^{2}$ =   2.3($\gamma$ - 1.83) 
+ (122.3 - 59.3 $\gamma$)c \\ \hline

a = 3/7  &   $\omega^{2}R^{2} $=   0.57($\gamma$ -    1.93) +
(15.2 - 5.1$ \gamma$)c  \\ \hline

a = 3.4/7 & $\omega^{2}R^{2}$ = 0.4($\gamma$ - 2.6)
+(8.9 - 2.3 $\gamma$)c \\  \hline
a = 3.49/7  &   $\omega^{2}R^{2}$ =   0.36($\gamma$ - 2.76) 
+ (8.0 -1.97 $\gamma$)c  \\ \hline

\end{tabular}
\caption{ $\omega^{2}$ $vs.$   $c$  for given values of $a$.}

\end{center}
\end{table}

\begin{table}
\begin{center}
\begin{tabular}{|l|l|l|r|}  \hline

a =  2/9  & $c_{max}$ = 0.0016 &$\gamma_{c}$ = 1.79 -6.87 c \\  \hline

a =  2/7  & $c_{max}$ = 0.0028 &$\gamma_{c}$ = 1.83 - 13.39 c \\  \hline

a =  3/7  & $c_{max}$ = 0.083 &$\gamma_{c}$ = 1.93 - 5.55 c \\  \hline

a =  3.4/7  & $c_{max}$ = 0.11 &$\gamma_{c}$ =  2.6 - 2.84 c \\  \hline

a =  3.47/7  & $c_{max}$ = 0.12 &$\gamma_{c}$ = 2.75 - 7.29 c \\  \hline

\end{tabular}
\caption{$\gamma_{c}$ $vs$   $c$  for given values of $a$.}

\end{center}
\end{table}


\section{CONCLUSIONS}

I have presented a summary of results obtained with Krsna Dev
concerning the existence of self-gravitating spheres in General
Relativity with anisotropic equations of state, aka {\it anisotropic
stars}. I have also presented a summary of our investigation of their 
stability, based on the extension of Chandrasekhar's celebrated
variational formalism for isotropic spheres to those with anisotropic
energy-momentum tensors.

Although the search for exact solutions restricts the forms of
anisotropy we could treat, our results illustrate the fact
that, indeed, pressure anisotropy may greatly affect the
physical structure of the star, leading to several
observational effects. Most importantly, the absolute stability bound
$2M/R < 8/9$ can be violated, and the star's surface redshift
may be arbitrarily large ($z_s > 2$). 
It is thus conceivable that objects which are
observed at large redshift may actually be closer than we think
due to anisotropic distortions. They may also be more stable then we think,
especially if pressure anisotropy exists near the stellar core.

Here, I have presented results only for one of the cases we treated, stars with
energy density scaling as $1/r^2$. These are of interest as they include
ultra-relativistic equations of state used to model the cores of neutron
stars. Perhaps the most important lesson of our study is that one must
keep an open mind as to whether isotropy is or not a justified
assumption to describe stellar matter. Until we have a better 
microscopic understanding of what truly goes on inside ultra-dense
compact objects, isotropy should be taken with a grain of salt. Especially
if some of these objects contain bosonic condensates at their cores, 
as is the case for several models of neutron star interiors and for 
a whole class of fully-anisotropic
hypothetical objects known as boson stars\cite{B-STARS}.

\acknowledgements

I'd like to express my gratitude to the organizers of IWARA 2003, 
and in particular to Helio Coelho, Sandra Prado, and Cesar Vasconcellos for 
their invitation and wonderful hospitality. My research is sponsored
in part by a NSF grant PHY-0099543.


\begin{thebibliography}{99}

\bibitem{CLAYTON} D. D. Clayton, {\it Principles of Stellar Evolution and 
Nucleosynthesis}, (The University of Chicago Press, Chicago, 1983);
R. Kippenhahn and A. Weigert, {\it Stellar Structure and 
Evolution}, (Springer-Verlag, Berlin, 1991).

\bibitem{GLENDENNING} N. K. Glendenning,  {\it Compact Stars: Nuclear Physics,
Particle Physics and General Relativity}, (Springer-Verlag, Berlin, 1997); H.
Heiselberg and M. H.-Jensen, Phys. Rep. {\bf 328}, 237 (2000).

\bibitem{RUDERMAN} M. Ruderman, Annu. Rev. Astron. Astrophys.,
 {\bf 10}, 427 (1972);
V. Canuto, Annu. Rev. Astron. Astrophys., {\bf 12}, 167 
(1974).

\bibitem{B-STARS} M. Gleiser, Phys. Rev. {\bf D38}, 2376 (1988);
M. Gleiser and R. Watkins, Nucl. Phys. {\bf B319}, 733 (1989).
For comprehensive reviews see,
A. R. Liddle and M. S. Marsden, Int. J. Mod. Phys.
{\bf D1} 101, (1992); P. Jetzer, Phys. Rep. {\bf 220}, 163 (1992); E. W.
Mielke and F. E. Schunck, in {\it Proceedings of 
8th M. Grossmann Meeting}, T. Piran (ed.), (World Scientific, Singapore, 1998).

\bibitem{SAWYER} R. Sawyer and D. Scalapino Phys. Rev. {\bf D7}, 382 (1973).

\bibitem{BOWERS}  R. L. Bowers and E. P. T. Liang, 
Ap. J. {\bf 188}, 
657 (1974);
J. Ponce de Leon, J. Math. Phys., {\bf 28},  1114 (1987);
M. Gokhroo and A. Mehra Gen. Rel. Grav., {\bf 26}, 75 (1994);
H. Bondi, Mon. Not. R. Astron. Soc. {\bf 259}, 365 (1992);
E. S. Corchero, Clas. Quantum Grav. {\bf 15}, 3645 (1998);
L. Herrera, Phys. Lett. {\bf A 165}, 206 (1992);
L. Herrera and N. O. Santos,  
Phys. Rep., {\bf 286}, 53, (1997).

\bibitem{kg1} K. Dev and M. Gleiser, Gen. Rel. Grav. {\bf 24}, 1793 (2002).

\bibitem{kg2} K. Dev and M. Gleiser, Gen. Rel. Grav. {\bf 35}, 1435 (2003).

\bibitem{ZAPOLSKY} C. Misner and  H. Zapolsky, Phys. Rev. Lett., {\bf 12}, 635
(1964).



\end{thebibliography}
\end{document}